\documentclass[AMA,STIX2COL,Linenumbersoff]{MRM}
\articletype{Research article Submitted to Magnetic Resonance in Medicine}%

\received{}
\revised{}
\accepted{}
\begin{document}

\title{Flexible metasurface for improving brain imaging at 7T}

\author[1]{Vladislav Koloskov*}{}

\author[2,3]{Wyger Brink}{}

\author[3]{Andrew Webb}{}

\author[1]{Alena Shchelokova}{}

\authormark{Vladislav Koloskov \textsc{et al}}

\address[1]{\orgdiv{School of Physics and Engineering}, \orgname{ITMO University}, \orgaddress{\state{St. Petersburg}, \country{Russian Federation}}}

\address[2]{\orgdiv{Magnetic Detection \& Imaging Group, TechMed Centre}, \orgname{University of Twente}, \orgaddress{\state{Enschede}, \country{The Netherlands}}}

\address[3]{\orgdiv{C.J. Gorter MRI Center, Department of Radiology}, \orgname{Leiden University Medical Center}, \orgaddress{\state{Leiden}, \country{The Netherlands}}}

\corres{*Vladislav Koloskov, ITMO University, Kronverkskiy Pr., 49A, St. Petersburg, 197101, Russia. \email{vladislav.koloskov@metalab.ifmo.ru}}

%\presentaddress{This is sample for present address text this is sample for present address text}

\finfo{The part of the work related to electromagnetic simulations and experimental characterization of the metasurface-based sheet was supported by the Russian Science Foundation (Project 21-19-00707). The part of the research devoted to the B$_1^+$ and SNR mapping was carried out with the support of the Ministry of Science and Higher Education of the Russian Federation (075-15-2021-592).}

\abstract[Summary]{
\section{Purpose} Ultra-high field MRI offers unprecedented detail for non-invasive visualization of the human brain. However, brain imaging is challenging at 7T due to the B$_1^+$ field inhomogeneity, which results in signal intensity drops in temporal lobes and a bright region in the brain center. This study aims to evaluate using a metasurface to improve brain imaging at 7T and simplify the investigative workflow.
\section{Methods} Two flexible metasurfaces, each comprising a periodic structure of copper strips and parallel-plate capacitive elements printed on an ultra-thin substrate, were optimized for brain imaging and implemented via PCB. We considered two setups: (1) two metasurfaces located near the temporal lobes; and (2) one metasurface placed near the occipital lobe The effect of metasurface placement on the transmit efficiency and specific absorption rate was evaluated via electromagnetic simulation studies with voxelized models. In addition, their impact on SNR and diagnostic image quality was evaluated in vivo for male and female volunteers.
\section{Results} Placement of metasurfaces near the regions of interest led to an increase in homogeneity of the transmit field by 5$\%$ and 10.5$\%$ in the right temporal lobe and occipital lobe for a male subject, respectively. SAR values changed insignificantly and were under recommended limits. In vivo studies also confirmed the numerically predicted improvement in field distribution and receive sensitivity in the desired ROI.
\section{Conclusion} Optimized metasurfaces enable homogenizing transmit field distribution in the brain at 7T. The proposed lightweight and flexible structure has the potential to provide MR examination with higher diagnostic value images.}

\keywords{7T MRI, brain imaging, B$_1^+$ inhomogeneity, metasurface, passive shimming}

\maketitle

% \footnotetext{Submitted to Magnetic Resonance in Medicine}

%Manuscript words: count

\clearpage

\section{Introduction}

Ultra-high field magnetic resonance imaging (MRI) is an advanced, non-invasive method for visualization of the human brain, offering unprecedented detailed imaging of anatomic and vascular structures~\cite{kraff20177t, balchandani2015ultra}. Compared with clinical MR machines with static field strengths (B$_0$) of 1.5T or 3T, ultra-high field 7T (and above) systems provide an increased signal-to-noise ratio (SNR)~\cite{duyn2012future, uugurbil2003ultrahigh, collins2001signal}, enabling higher-resolution brain imaging, as well as an increase in sensitivity of susceptibility-weighted imaging. Thus, 7T MRI allows one to visualize the brain's anatomical structures with improved detection of lesions~\cite{shaffer2022ultra, noebauer2015brain}, more accurate measurement of diffusion~\cite{kraff20177t} and blood flow~\cite{teeuwisse2010arterial, park2018advances} and more precise functional mapping ~\cite{viessmann2021high, bubrick20227t, trattnig2018key}. Nevertheless, the increase of the static field strength (i.e., operating frequency) leads to challenges, such as the shortening of the radiofrequency (RF) wavelength in tissue and the appearance of standing waves in the RF field distribution~\cite{collins2005central, webb2010parallel, karamat2016opportunities}. When imaging the head at 7T, this leads to a pronounced inhomogeneity of the transmit radiofrequency (RF) field -- with a maximum B$_1^+$ field in the brain's center and pronounced intensity drops near the temporal lobes.

One of the ways to improve the B$_1^+$ uniformity at 7T is active RF-shimming, where an array of RF coils are individually driven with specific phases and/or magnitudes~\cite{vandamme2021universal, deniz2019parallel}. Parallel transmission has become helpful for biomedical research; however, its application in clinics is limited by the hardware complexity and difficulties in ensuring compliance with RF safety limits~\cite{fiedler2020safety, mcelcheran2019numerical}. Alternatively, passive RF shimming methods utilizing materials such as high permittivity dielectric pads and artificial metal-dielectric structures have been proposed~\cite{webb2022novel}. To date, dielectric pads remain the only clinically available way to improve the B$_1^+$ uniformity. In particular, high-permittivity dielectric pads are relatively inexpensive and offer a robust method to enhance the quality of image in the brain~\cite{teeuwisse2012simulations, brink2014high} without increasing the specific absorption rate (SAR)~\cite{webb_dielectric_2011, brink2023radiofrequency}. It was demonstrated that flexible dielectric pads based on a suspension of calcium or barium titanate in deuterated water could improve the homogeneity of the B$_1^+$ field across the brain at 7T~\cite{teeuwisse2012simulations}, or in a local region-of-interest (ROI) enabling high spatial resolution imaging of the inner ear~\cite{brink2014high}, enlarge the field of view of receive coil arrays and boost the accuracy of functional MRI studies~\cite{vaidya_improved_2018}.

Nonetheless, dielectric pads need tailored geometrical dimensions and a 5-15 mm thickness to obtain the desired effect~\cite{van2019high}. As a result, the space required between the subject and the receive coil can be limiting and may impact the patient's comfort. Also, they have limited shelf life as the materials dry out, i.e., the dielectric properties of the pad change over time~\cite{o2016practical}.

Another passive shimming approach uses materials with artificial properties that can not be obtained in nature –- metasurfaces -- compact two-dimensional periodic metal-dielectric structures. The properties of metasurfaces can be easily optimized and configured by changing their design~\cite{vorobyev2020artificial, alipour2021improvement, sokol2022flexible, gomez2022hilbert} and offer several advantages as an alternative to dielectric pads. For example, previous studies have shown that for abdominal imaging at 3T, dielectric pads can be replaced by a compact and flexible metasurface to modify the B$_1^+$ distribution and improve artifacts associated with RF non-uniformity~\cite{vorobyev_improving_2022}. The proposed metasurface was a two-dimensional periodic structure made of metal planar crosses connected with parallel-plate capacitors printed on the dielectric substrate, allowing the RF field to be tailored effectively. The advantages of metasurfaces over conventional dielectric pads are compactness and low weight, involving a flexible sheet of less than 1 mm thickness easily placed inside a close-fitting RF coil array. Metasurfaces also offer long-term durability, unlike aqueous suspensions that dry out over time, which reduces their permittivity. In addition, the properties of the metasurfaces dictated by their geometry (design) can be easily varied depending on the specific imaging goal or application area.

This work aims to evaluate the impact of a flexible metasurface on brain visualization at 7T MRI. Several metasurfaces were designed and optimized for imaging the temporal lobes and the occipital part using electromagnetic simulations and manufactured using flexible printed circuit board (PCB) technology. The impact of the metasurfaces on the B$_1^+$ field distribution in the desired ROI and SAR was evaluated through numerical analysis. Experimental studies with two healthy volunteers (male and female) have been performed to estimate the qualitative and quantitative effects of the metasurfaces when placed in a commercial head coil in terms of transmit efficiency, SNR, and intercoil coupling in the RF receive array.

\begin{figure*}[htbp]
	\centering 
	\includegraphics[width=0.9\linewidth]{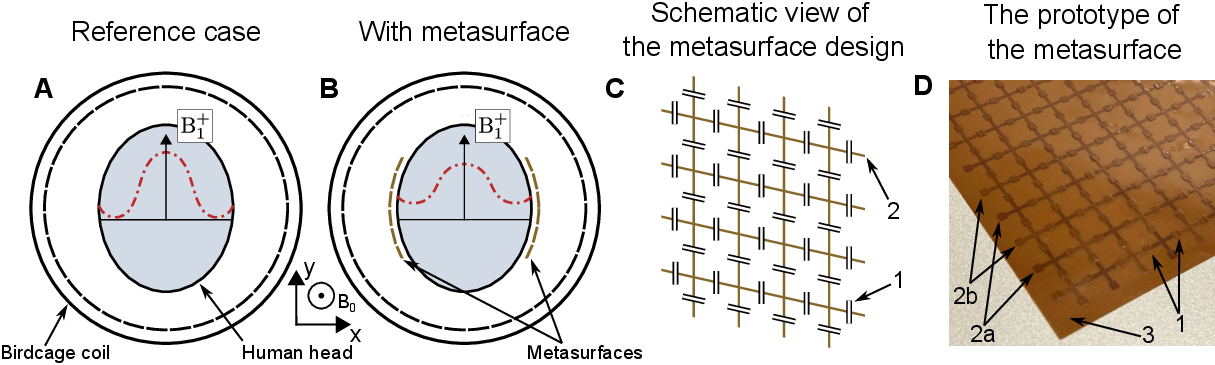}
	\caption{Schematic view of the proposed metasurface design and its operation principle inside a head coil. A, Schematic view of the birdcage-type head coil and a head phantom. A dielectric-induced standing wave artifact: low B$_1^+$ amplitude in the temporal lobes and high amplitude of the B$_1^+$ field in the brain's center. B, The same setup, but with the metasurfaces on the left and right from the phantom, providing the additional magnetic field helping to increase the B$_1^+$ field amplitude at the side parts of a brain. C, A schematic representation of the metasurface design based on unit cells made of metal strips (2) connected with capacitors (1). D, A prototype of the proposed metasurface based on a dual-sided printed circuit board. The top (2b) and bottom (2a) layers of metallization are separated by a dielectric substrate (3).}
	\label{Fig1}  
\end{figure*}

\section{Methods}

\subsection{Electromagnetic simulations}

All electromagnetic simulations were performed in CST Studio Suite 2021 (Dassault Systèmes, France). These calculations aimed to find the optimal geometry of the metasurfaces for the brain imaging task and estimate their impact on the transmit RF field and ${\text{SAR}_{\text{av.10g}}}$. The B$_1^+$ field maps and ${\text{SAR}_{\text{av.10g}}}$ were normalized to 1~W of total accepted power. The numerical setups are schematically shown in Figs.~\ref{Fig1}(A,B). As a transmit coil, we used a 16-leg high-pass birdcage head coil with a 200~mm length and 290~mm diameter with two quadrature excited feeds tuned and matched to 50 Ohms at 300~MHz. The birdcage coil was loaded with the voxelized male (Hugo) and then with the female (Ella) human model with a voxel resolution of ${5}\times{5}\times{5}$~mm$^{3}$ from the Voxel Family (IT’IS Foundation). Figure~\ref{Fig1}(A) shows the reference case (without the metasurface) when the amplitude of the B$_1^+$ field drops at the side parts of the head (near the temporal lobes) due to the ``standing wave'' artifact. With the addition of the metasurface, the incident electromagnetic field excites conductive currents flowing via a metal grid, resulting in modified field distribution, i.e., the secondary magnetic field of the metasurface compensates the B$_1^+$ intensity voids (Fig.~\ref{Fig1}(B)). 

The proposed metasurface was modeled as a periodic structure of metal crosses (${10}\times{10}$~mm$^{2}$) connected by capacitors with variable values of capacitance (Fig.~\ref{Fig1}(C)). Previously, a similar design was proven feasible for abdominal imaging at 3T, providing the same B$_1^+$ homogeneity improvement as a conventional dielectric pad~\cite{vorobyev_improving_2022}. The number of unit cells, their size, and the value of capacitors in the metasurface design will affect the equivalent electromagnetic properties, current distribution, and secondary magnetic field profile.

Thus, the metasurface design procedure was set up by defining the capacitance value to improve the B$_1^+$ field distribution in the regions of interest (ROIs). The overall dimensions of the metasurfaces were chosen as ${180}\times{180}$~mm$^{2}$ similar to the dielectric pads, which were shown optimal for B$_1^+$ field homogenization in the brain area at 7T~\cite{teeuwisse2012simulations}. During the optimization, we fixed the period of the structure (d=11 mm) and the number of unit cells (N=15). First, we varied the capacitance between 4 and 12~pF in steps of $1$~pF. Then, we fine-tuned the value in smaller steps of $0.1$~pF to achieve a gain in $|\text{B}_1^+|$ amplitude and reduce its inhomogeneity (coefficient of variation) in the ROI. The coefficient of variation (Cv) of $|\text{B}_1^+|$ was calculated as the ratio of a standard deviation of the field and the mean value in the desired ROI. 

The two-ports circularly-polarized head birdcage coil at 7T MRI naturally produces a certain degree of left-right asymmetry in the B$_1^+$ distribution~\cite{brink2018simple, brink2014high, teeuwisse2012simulations}. In other words, a more pronounced minimum from one side of the brain occurs, especially in subjects with larger heads. Therefore, we performed the optimization of the metasurface geometry for temporal lobes, considering two configurations: one with two identical metasurfaces (similar dimensions, number of unit cells, and capacitance) and one with two different metasurfaces (similar dimensions and number of unit cells, but different capacitance). 

Once the best geometries for temporal and occipital lobes were selected, we evaluated the effect of the metasurfaces on the local SAR averaged over 10 grams of tissue (${\text{SAR}_{\text{av.10g}}}$) required for safety assessment in different configurations without and with artificial structures. 

\subsection{Metasurface manufacturing}

Each metasurface was manufactured as a flexible PCB with two metallization layers separated from each other by a flexible dielectric substrate (Fig.~\ref{Fig1}(D)), with dimensions of ${180}\times{180}\times{0.025}$~mm$^{3}$ with a relative permittivity value of $\varepsilon_r = 3.4$ (DuPont$\textsuperscript{\texttrademark}$ Pyralyx$\textsuperscript{\textregistered}$ AP8515R). To make the structure compact, numerically modeled lumped elements capacitors were replaced by rectangular patched capacitors according to the formula $C=\varepsilon_0 \varepsilon_r S/d$, where $\varepsilon_r$ and the thickness $d$ were given from the substrate. To obtain capacitors with 4.8~pF, patches with dimensions of ${2}\times{2}$~mm$^{2}$ were used, as for the 7~pF case -- patches with dimensions of ${2.4}\times{2.4}$~mm$^{2}$. To protect the metasurface, we placed the structure inside a case made of leather to maintain the structure's flexibility.

\subsection{MR imaging}

All experimental studies with volunteers were performed on a Philips 7T Achieva system (Philips Healthcare, Best, The Netherlands) with a quadrature transmit head coil (NM-008A-7P, Nova Medical, Wilmington, MA) and a 32-channel receive phased array (NMSC025-32-7P, Nova Medical, Wilmington, MA). One female (age 32) and two male (age 27 and 59) volunteers were studied under a protocol approved by the local medical ethics board. Three configurations were studied: (1) reference case without any metasurfaces, (2) with the metasurface placed posterior, and (3) with two metasurfaces placed near temporal lobes. Also, for the male volunteer (age 59), we performed additional studies with two dielectric pads placed near the temporal lobes to compare the proposed approach with the state-of-the-art one. B$_1^+$ maps were generated using the dual refocusing echo acquisition mode (DREAM) sequence with the following parameters: field of view = $256\times256\times192$~mm$^3$, in-plane spatial resolution = $4\times4$~mm$^2$, slice thickness = 4~mm, 48 slices, TR/TE = 4/2.38~ms, STEAM/imaging tip angle = 50$^{\circ}$/10$^{\circ}$. SNR maps were generated using a 3D spoiled GRE sequence with the following parameters: field of view = $256\times256\times192$~mm$^3$, in-plane spatial resolution = $2\times2\times2$~mm$^3$, TR/TE = 3.32/1.97~ms, turbo factor = 900, non-selective excitation, tip angle = 1$^{\circ}$. Anatomical images were acquired using a T$_1$-weighted MP-RAGE sequence with the following parameters: field of view = $256\times256\times192$~mm$^3$, in-plane spatial resolution = $1\times1\times1$~mm$^3$, TR/TE/TI = 4.94/2.34/1050~ms, turbo factor = 256, shot interval = 2500 ms, 69 shots, tip angle = 5$^{\circ}$). All post-processing of experimental data, including B$_1^+$ and SNR maps as well as noise correlation matrices, was done using MATLAB R2022b (The Mathworks, Natick, MA, USA).

To address subject movement between acquisitions with and without metasurfaces, co-registration and normalization of the SNR maps were performed to address subject movement between acquisitions with and without metasurfaces. All maps acquired with metasurfaces were normalized to the reference case, i.e., without metasurfaces, to quantify the local SNR gain. This was done using the Statistical Parametric Mapping (The Wellcome Centre for Human Neuroimaging, UCL Queen Square Institute of Neurology, London, UK) MATLAB-based toolbox~\cite{penny2011statistical}.

\section{Results}

\subsection{Electromagnetic simulations}

As a result of the optimization process with the male voxel model, two configurations of the metasurfaces were implemented: one with 4.8~pF and one with 7~pF capacitors in the structure. Initial results showed that having two different metasurfaces on either side of the head provided better results regarding B$_1^+$ field uniformity than the placement of two identical metasurfaces. This setup with two lateral metasurfaces will be further denoted as ``two MSs''. Additionally, a single metasurface with 4.8~pF capacitors was optimal for improving the B$_1^+$ field distribution in the occipital lobe region, which will be further denoted as ``one MS''. The same configurations were evaluated in the female voxel model.

\begin{figure}[!h]
    \centering
	\includegraphics[width=0.9\linewidth]{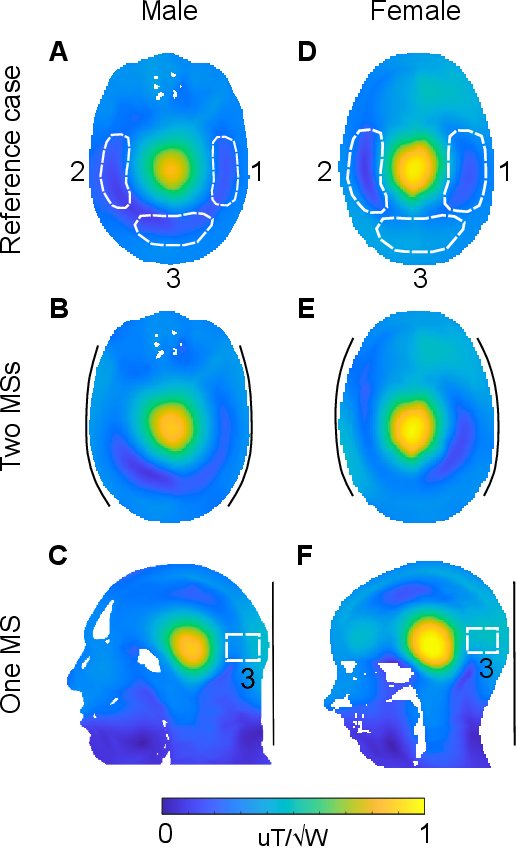}
	\caption{Numerically calculated ${|\text{B}_1^+|}$ maps for male (left column) and female (right column) voxel models inside the head birdcage coil: A, D, without and B, E, with two metasurfaces located near temporal lobes, and C, F, with one posterior placed metasurface, correspondingly. B$_1^+$ maps were normalized to 1~W of total accepted power. Numbered white dashed lines represent ROI$_1$, ROI$_2$, and ROI$_3$ where the ${|\text{B}_1^+|}$ and its coefficient variation (Cv) were evaluated. Black lines schematically represent the position of the metasurfaces.}
	\label{Fig2}  
\end{figure}

Figure~\ref{Fig2} shows the numerically evaluated $|\text{B}_1^+|$ maps in the male and female voxel models' central axial and sagittal slices. One can observe that the metasurfaces improved the areas of low $|\text{B}_1^+|$. For the male voxel model, it was improved from $0.363$~uT by $2.4\%$ and $3.3\%$ for the cases with one and two metasurfaces, respectively. Whereas for the female voxel model, the mean B$_1^+$ value was improved by $2.3\%$ for the case with one posterior placed metasurface compared to the reference case with $0.426$~uT. As for the case with two metasurfaces, no positive changes were present.

\begin{figure}[!h]
    \centering
	\includegraphics[width=0.9\linewidth]{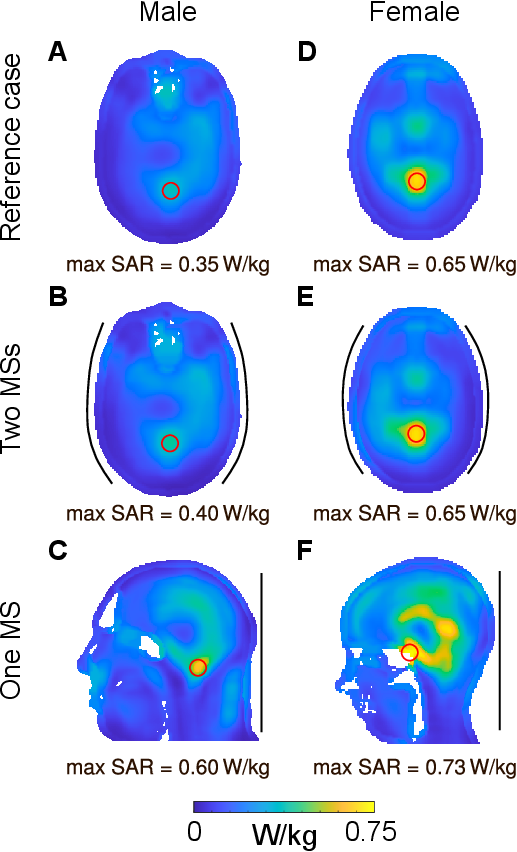}
	\caption{Numerically calculated SAR$_{\text{av.10g}}$ maps for male (left column) and female (right column) voxel models inside the head birdcage coil: A, D, without,  B, E, with two metasurfaces located near temporal lobes, and C, F, with one posterior placed metasurface, correspondingly. For each case, maximum local SAR values were calculated in the slice crossing the center of the metasurfaces with $z=0$ or $x=0$. Hot spots with maximum local SAR values within the corresponding slice are encircled in red for each numerical setup. SAR maps were normalized to 1~W of total accepted power. Black lines schematically represent the position of the metasurfaces.}
	\label{Fig3}  
\end{figure}

Table~\ref{Tab_1} shows the quantitative results of electromagnetic simulations for male and female voxel models. $|\text{B}_1^+|$-values were calculated over the whole brain region. Cv-values were calculated in regions of interest with 1~cm slice thickness depicted by white dashed lines representing right and left temporal lobes (ROI$_1$ and ROI$_2$, correspondingly) and occipital lobe (ROI$_3$). Note that for male and female voxel models with posterior metasurface, the B$_1^+$ field homogeneity was improved in ROI$_3$ by $10.5\%$ and $1\%$, respectively. Whereas, for two metasurfaces, it was improved by $5\%$ and $11.4\%$ in ROI$_2$ where the reference $|\text{B}_1^+|$ was lowest.

\begin{table*}[h!]
\caption{Numerically calculated results of transmit and SAR characteristics for the male and female voxel models. ${\text{B}_1^+}$ values were calculated in the whole brain region. Cv values were calculated in corresponding ROIs with 1~cm slice thickness. ROI$_1$ -- right temporal lobe, ROI$_2$ -- left temporal lobe, ROI$_3$ -- occipital lobe. SAR values were calculated for 1~W of total accepted power. Maximum local SAR values were evaluated for axial slice $z=0$.}
\begin{center}
\begin{tabular}{lccccc}
\hline
\multicolumn{1}{c}{\multirow{2}{*}{}} & \multirow{2}{*}{${\text{B}_1^+}$} & \multicolumn{3}{c}{Cv}   & \multirow{2}{*}{${\text{max SAR}}_{\text{av.10g}}$} \\ \cline{3-5}
\multicolumn{1}{c}{}                  &                          & ROI$_1$   & ROI$_2$   & ROI$_3$   &                          \\ \hline
Male                                  &                          &        &        &        &                          \\
Reference case                        & 0.363 uT                 & 13.5\% & 21.3\% & 24.7\% & 0.52 W/kg                \\
One metasurface                       & 0.372 uT                 & 14.4\% & 18.3\% & 14.2\% & 0.65 W/kg                \\
Two metasurfaces                      & 0.375 uT                 & 12.7\% & 16.3\% & 26.9\% & 0.63 W/kg                \\
Female                                &                          &        &        &        &                          \\
Reference case                        & 0.426 uT                 & 18.1\% & 32.7\% & 10.4\% & 0.71 W/kg                \\
One metasurface                       & 0.436 uT                 & 26.5\% & 32.0\% & 9.5\%  & 0.73 W/kg                \\
Two metasurfaces                      & 0.420 uT                 & 25.4\% & 21.3\% & 12.0\% & 0.73 W/kg                \\   \hline
\end{tabular}
\end{center}
\label{Tab_1}
\end{table*}

The numerically evaluated ${\text{SAR}_{\text{av.10g}}}$ distribution in the central axial and sagittal slices ($z=0$ and $x=0$, respectively) for all cases (Figure~\ref{Fig3}) show that no significant changes were present. Maximum-value locations were in other axial slices with unchanged values under recommended limits for both voxel models and all configurations. Note that $\text{max SAR}_{\text{av.10g}}$ values evaluated in the whole head volume are presented for all cases in Table~\ref{Tab_1}. 

\subsection{MR imaging}

\begin{figure*}[]
    \centering
	\includegraphics[width=\linewidth]{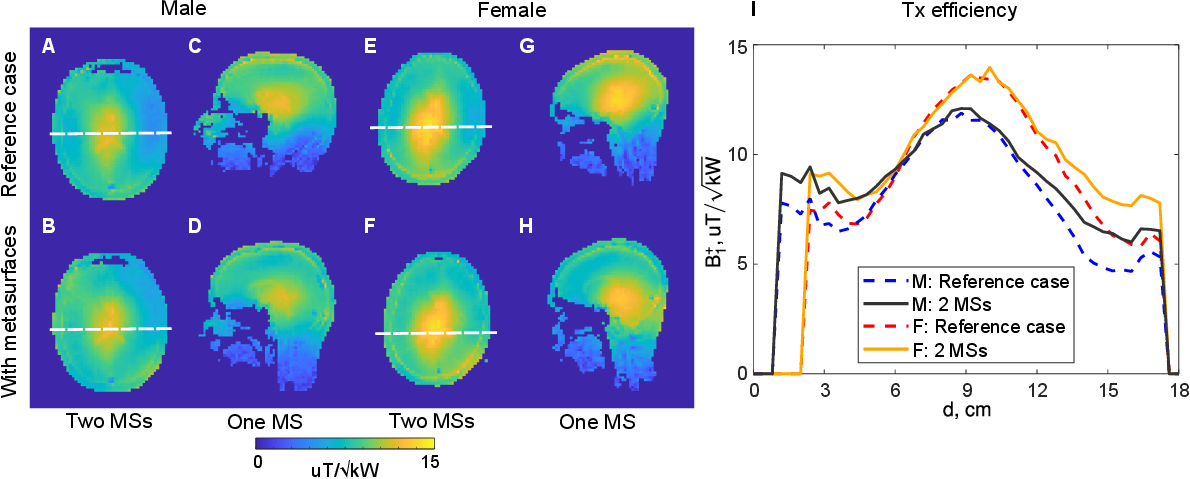}
	\caption{Measured transmit efficiency improvement for male and female subjects. Transverse cross-sections of the transmit efficiency without (A,E) and with (B,F) two metasurfaces located near temporal lobes. Sagittal cross sections of the brain without (C,G) and with (D,H) one posterior placed metasurface, correspondingly. I, Measured transmit efficiency in the subjects along white dashed lines depicted in panels A, B and E, F, for male (M) and female (F) subjects, respectively. The distance (d) is measured from the left end of the white dashed line (0-value) in the left-right direction. The transmit efficiency maps show the effectiveness of the metasurfaces for locally tailoring the B$_1^+$ field.}
	\label{Fig4}  
\end{figure*}

Figure~\ref{Fig4} demonstrates the measured effect of the metasurfaces on the B$_1^+$ field uniformity and efficiency for both male and female volunteers. The transverse and sagittal cross-sections of B$_1^+$ field maps are presented for the reference case Fig.~\ref{Fig4}(A,E and D,G), with two metasurfaces near both temporal lobes Fig.~\ref{Fig4}(B,F) and one posterior metasurface Fig.~\ref{Fig4}(D,H). Worthy of note is that the dimensions of the subject of study can vary from one patient to another, so there are slightly different reference field patterns. However, metasurfaces work almost identically in both cases. In other words, redistribution of B$_1^+$ can be seen in the same areas but with slightly varying amplitude.

To evaluate the effect of the proposed metasurface on the B$_1^+$, we acquired experimental data to qualitatively and quantitatively estimate the impact of dielectric pads and metasurfaces on a healthy male volunteer. Transverse cross-sections of B$_1^+$ maps can be viewed in Supplementary Fig.~\ref{FigS1} (A-C). One can observe that the low transmission efficiency in the left and right temporal lobes is improved with both dielectric pads and metasurfaces. Fig.~\ref{FigS1} (D) shows estimated values of the transmit field along the dashed lines in the left-right direction. Cases with dielectric pads and metasurfaces show the same effect in tailoring RF magnetic field distribution in ROIs.

\begin{figure*}[]
    \centering
	\includegraphics[width=\linewidth]{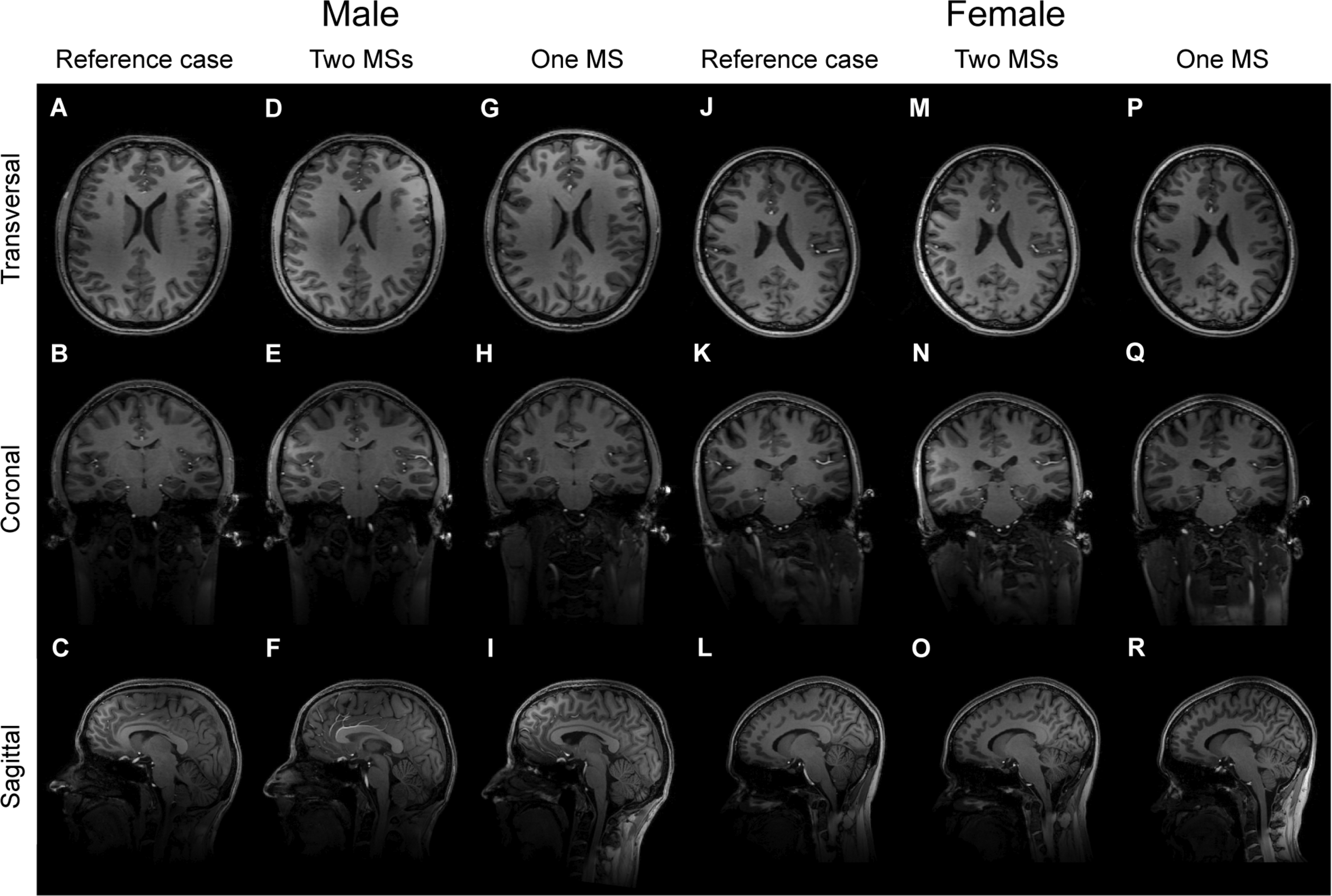}
	\caption{T$_1$-weighted images from healthy male and female volunteers for the reference case (without any metasurface, panels A-C and J-L), with two lateral metasurfaces (D-F, M-O), and with one posterior placed metasurface (G-I, P-R).}
	\label{Fig5}  
\end{figure*}

Experimentally obtained T$_1$-weighted MR images in the central transversal (first row), coronal (second row), and sagittal (third row) planes of the brain for the male and female volunteers are presented in Figure~\ref{Fig5}. For both volunteers, three cases were investigated: the reference case (A-C and J-L), with two metasurfaces near both temporal lobes (D-F and M-O) and with one metasurface placed near the occipital lobe (G-I and P-R). One can notice that in the transversal and coronal planes for the case with two metasurfaces, the brain areas close to the metasurfaces are brighter compared to the reference case. For the metasurface placed at the posterior of the head, areas around the cerebellum and neck increase in signal intensity. 

\begin{figure*}[]
    \centering
	\includegraphics[width=\linewidth]{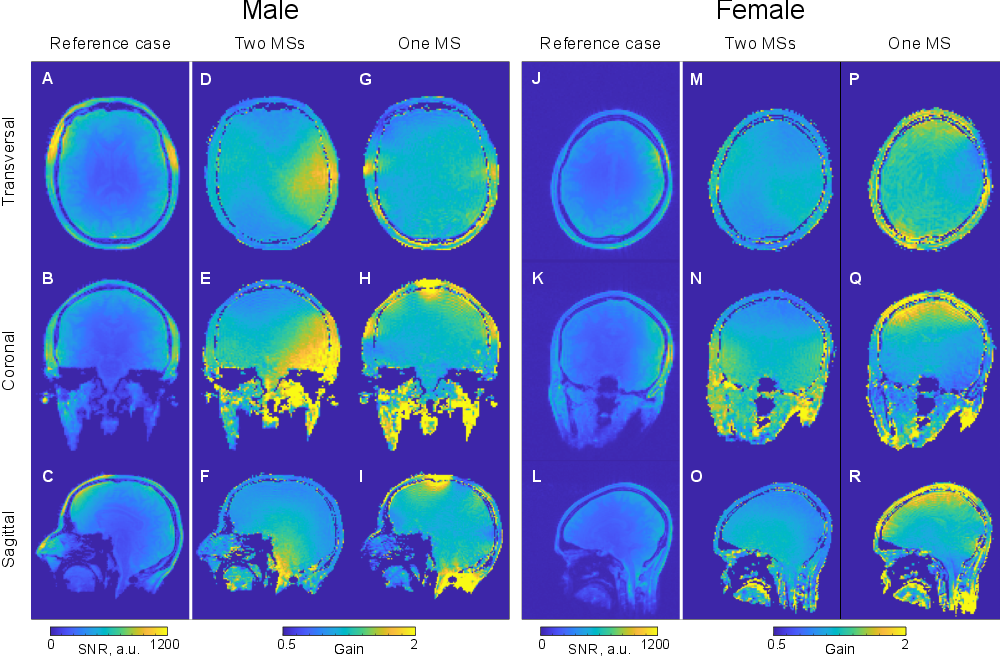}
	\caption{SNR and SNR gain maps for healthy male and female volunteers for the reference case (A-C, J-L), with two lateral metasurfaces (D-F, M-O), and with one posterior placed metasurface (G-I, P-R). Adding the metasurfaces increased The receive sensitivity for almost all brain volumes.}
	\label{Fig6}  
\end{figure*}

The experimentally measured SNR maps are presented in Figure~\ref{Fig6}. For both male and female subjects, metasurfaces placed near the regions of interest improved the Rx sensitivity not only in the cerebral cortex but also in the inner structures of the brain, with the values of SNR gain of more than 1. Note that areas with values above the color map limit in SNR gain maps around the subject occurred because of co-registration and normalization of the SNR maps, so they should not be considered.

The proposed metasurfaces are resonant at frequencies well away from the Larmor frequency ($300$~MHz). This was experimentally verified with an S$_{11}$ network analyzer measurement, which showed a peak centered at $445$~MHz for the metasurface with $4.8$~pF and at $368$~MHz for the metasurface with $7$~pF with full-width-half-maximum of ${\sim}20$~MHz. However, they are placed close to the receive coil array. Therefore, additional experimental studies were performed to evaluate their effect on inter-element decoupling by evaluating the noise correlation matrix. In the best case, no coupling should be present (and noise uncorrelated). A schematic representation of the channel layout of the 32-channel receive coil used in the experiment is presented in Figure~\ref{FigS2} (A). The noise correlation matrices without metasurfaces, with metasurfaces, and the respective difference are shown in supporting material Figure~\ref{FigS2} (B,E), (C,F), and (D,G), correspondingly. The maximum absolute difference value is not more than $0.28$ (for a female subject) and not more than $0.16$ (for a male subject) for the channels that are close to the metasurfaces: 2, 3, 19, 18, and 15, 14, 30, 31 for the case with two metasurfaces and 22, 23, 26, 27 for the case with one metasurface. For other channels, no significant changes for both experimental configurations were registered. Note that as NCMs and their difference matrices have more pronounced differences for the female volunteer, in Fig.~\ref{FigS2}, results only for that one are presented.

\section{Discussion and Conclusions}

This work evaluated the effect of metasurfaces on improving ultra-high brain imaging, allowing further clinical assessment of this method in diagnostic studies. The proposed approach using ultra-thin passive metal-dielectric structures allows correcting the B$_1^+$ field inhomogeneities in the temporal and occipital lobes, which makes them very easy to fit even within the close-fitting multichannel receive arrays that are typically used. This substantially improves practical utility compared to conventional dielectric pads made from ceramic powder, overcoming material limitations and increasing patient comfort.

Via numerical simulations, we obtained optimized configurations of the metasurfaces in a male head model. Due to the intrinsic left-right difference in B$_1^+$ present without metasurfaces, we used two metasurfaces with different designs (one with 4.8~pF and another with 7~pF) to reach a similar $|\text{B}_1^+|$ in both temporal lobes. The same setup was used to estimate the effect in a female head model. A less pronounced left-right difference in B$_1^+$ field distribution was present in this case. Also, one can observe less effect on the redistribution in the desired ROIs.

It is important to note that the left-right asymmetry in B$_1^+$ field does not only appear when using quadrature birdcage coils but can also be observed when using transmit arrays~\cite{lakshmanan2021improved,williams2021nested}. Consequently, using optimized metasurfaces on either side of the head offers an effective means of correcting this intrinsic asymmetry without requiring changes in the size.

When simulating the RF transmit field with the male and female voxel models, one can observe the redistributed B$_1^+$ field in the left and right motor cortex and the occipital lobe (Fig.~\ref{Fig2}(B,E) and Fig.~\ref{Fig2}(C,F)). For the male subject, higher mean B$_1^+$ and homogeneity values were obtained in the proximity of the metasurfaces compared to the setup without them. As for the female subject, only insignificant improvements were present except in ROI$_2$. The decrease in mean B$_1^+$ value case with two metasurfaces placed near the female voxel model is based on the redistribution of the field throughout the whole brain region (Fig.~\ref{Fig2} (E))~\cite{teeuwisse2012quantitative}. In the meantime, for both voxel models, only minor changes in SAR values were registered (Fig.~\ref{Fig3}, Table~\ref{Tab_1}) with values below the corresponding limits recommended by the International Electrotechnical Commission (3.2~W/kg for head SAR)~\cite{international2010international}.

In vivo results approve the ones evaluated numerically. Fig.~\ref{Fig4} (A-H) shows qualitative redistribution of measured B$_1^+$ maps for two volunteers, especially notable in the right motor cortex (ROI$_2$) and the occipital lobe (ROI$_3$). Fig.~\ref{Fig4}(I) quantitatively estimates the investigated effect. One can observe a slight difference in the left-right asymmetry in the B$_1^+$ distribution for male and female volunteers, with less pronounced for the last one. Also, it is worth noting that the proposed metasurface affects the B$_1^+$ mainly within its vicinity, with negligible effects in the brain's center. So, a more homogeneous distribution can be achieved throughout the whole brain, i.e., the difference in the RF field intensity between the central and the side parts of the brain can be reduced.

Moreover, additional changes in receive sensitivity can be detected with metasurfaces placed near the subject. In Fig.~\ref{Fig6}, one can find the improvement in SNR in the cerebral cortex, inner ear, and cerebellum for both volunteers. The noise correlation measurements (Fig.~\ref{FigS2}) show no systematic changes in coupling between elements of the receive array. Notwithstanding the size of the metasurface being comparable to the size of the head and receive coil, only minor positive and negative variations in noise correlation (less than $0.28$ in absolute value) were observed for the channels close to the artificial structures.

Improvements in transmit efficiency and receive sensitivity with the proposed metasurface-based sheet led to enhanced image quality, as seen in the T$_1$-weighted images presented in Fig.~\ref{Fig5}. Positioning the metasurface near the occipital lobe can effectively extend the field of view of the transmit volume coil towards the neck (Fig.~\ref{Fig5}(I,R)) and substantially improve cerebellum contrast, which is typically a challenge at 7T. As the proposed metasurface did not substantially affect the receive array noise correlation and improved the B$_1^+$ similarly to the conventional dielectric pads, it may be considered a sustainable alternative to the state-of-the-art approach. In particular, the proposed metasurface offers several practical advantages compared to conventional dielectric pads. The metasurface is light, flexible, and scalable, which can be tailored for different ROIs. Producing the suggested metasurface is cost-effective and convenient as it involves printing a design on a PCB, which is an easily accessible and affordable manufacturing service that also ensures a longer lifespan of the structure while remaining operational.

This study has some limitations. The first and foremost is related to the positioning of the metasurfaces during numerical simulations, as the simulation grid does not allow for a consistent representation of a curved metasurface, which is how it is positioned during the experiment. This may affect the simulated B$_1^+$ field gains, which might be underestimated compared to experimental results. Another challenge is related to the speed of the optimization process, as each new configuration of the metasurface should be simulated separately. Dedicated optimization tools, as developed previously for dielectric pads~\cite{van2019high}, would shorten the design process and could lead to improved application-specific metasurface designs.

In conclusion, this study shows that challenges related to inhomogeneities of the RF transmit field in human brain imaging at 7T can be addressed using metasurfaces. This passive shimming approach might be beneficial for diagnosing brain disorders in functional brain MRI and brain MR elastography~\cite{murphy2019mr} without any changes in hardware, MR protocol, and duration of the MR procedure.

\section*{Acknowledgements}
The part of the work related to electromagnetic simulations and experimental characterization of the metasurface-based sheet was supported by the Russian Science Foundation (Project 21-19-00707). The part of the research devoted to the B$_1^+$ and SNR mapping was carried out with the support of the Ministry of Science and Higher Education of the Russian Federation (075-15-2021-592).

\subsection*{Conflict of interest}

The authors declare no potential conflict of interests.

\bibliography{main}

%\vfill\pagebreak

%\clearpage
%\section*{Supporting information}
\setcounter{figure}{0} 

\begin{figure*}[]
    \centering
	\includegraphics[width=\linewidth]{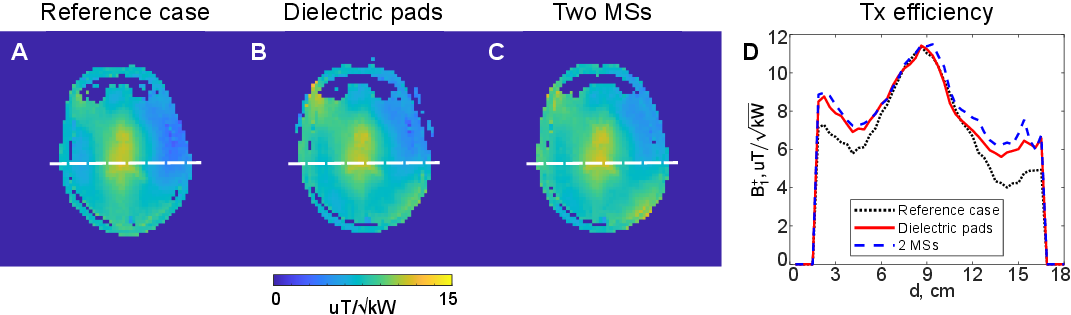}
        \renewcommand{\thefigure}{S\arabic{figure}}
	\caption{Transverse cross sections of the measured transmit efficiency in a male volunteer A, for the reference case, B, with dielectric pads, and C, with two metasurfaces. D, measured Tx efficiency in the subject along white dashed lines depicted on A,B, and C. The distance (d) is measured from the left end of the white dashed line (0-value) in the left-right direction.}
	\label{FigS1}  
\end{figure*}

\begin{figure*}[]
    \centering
	\includegraphics[width=\linewidth]{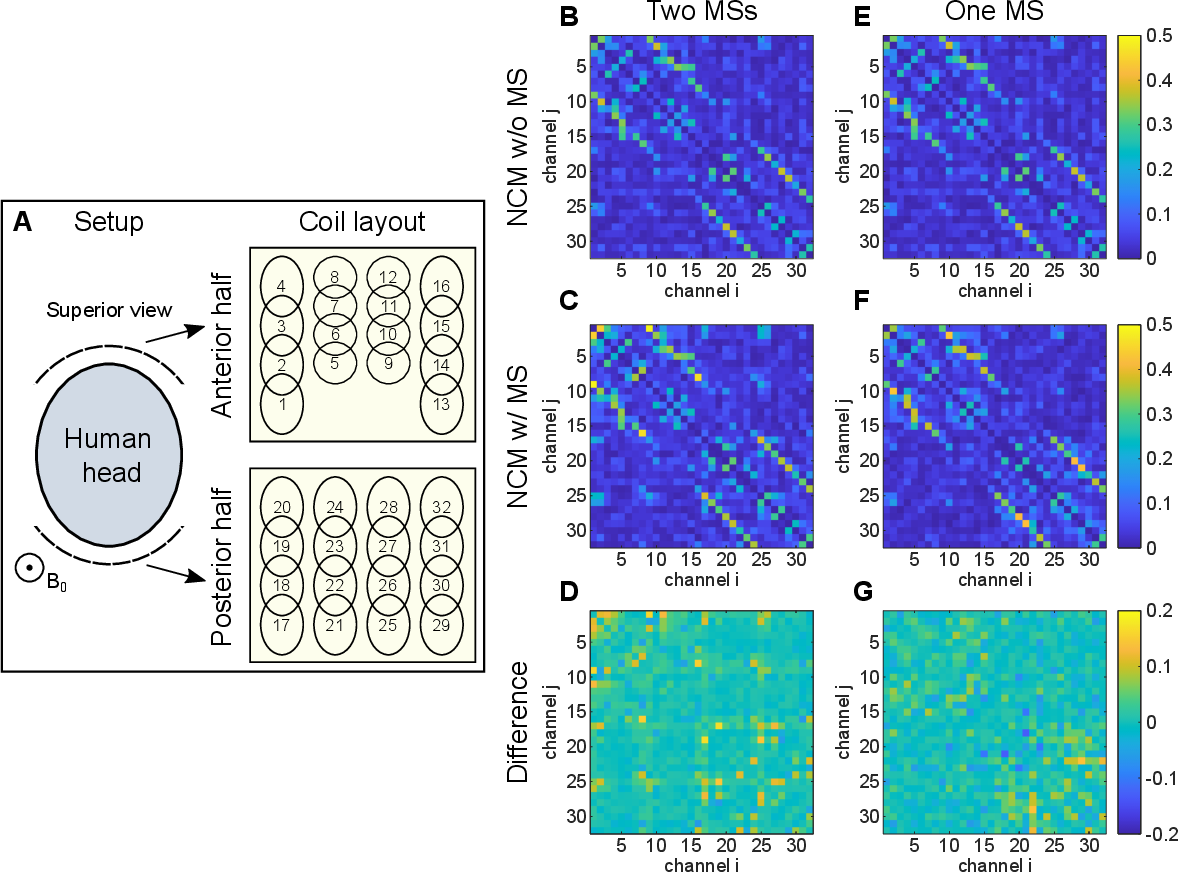}
        \renewcommand{\thefigure}{S\arabic{figure}}
	\caption{A, A general view of the setup and 32-channels receive coil layout; the noise correlation matrix for B and E, the reference case; for C, two metasurfaces placed near the temporal lobes, and F, one metasurface placed posterior; the difference between reference case and the noise correlation matrix for D, two metasurfaces, and G, for posterior metasurface. The noise correlation and differences matrices are related to the female volunteer.}
	\label{FigS2}  
\end{figure*}

\end{document}